# Nirjas: An open source framework for extracting metadata from the source code

Ayush Bhardwaj
Department of Computer Science & Engineering, ASET
Amity University, Noida, India
classicayush@gmail.com

Sahil
Department of Computer Science & Engineering, ASET
Amity University, Noida, India
sjha200000@gmail.com

Kaushlendra Pratap
Department of Computer Science & Engineering, ASET
Amity University, Noida, India
kaushlendrapratap.9837@gmail.com

Gaurav Mishra
Sathyabama University, India
gmishx@gmail.com

*Abstract*— Metadata/Comments are the critical element of any software development process. In this paper, we went on explaining how the metadata/comments in the source code can play an essential role in comprehending the software. We introduced a python based open-source framework "Nirjas" that helps us in extracting these metadata in a structured manner. There are various syntax, types and widely accepted conventions for adding a comment in the source file of different programming languages. Various edge cases can create noise in our extraction, for which we used Regex to accurately retrieve these metadata. The non-Regex method can give us the result but misses out on accuracy and noise separation. Nirjas can also separate different types of comments, source code and provide us with the details about those comments in terms of the line number, file name, language used, total SLOC, etc. Nirjas is a standalone python framework/library and can be easily and quickly installed using the source installation or using pip (package installer for Python). Nirjas was first created and started out as one of the projects during Google Summer of Code program and is currently developed and maintained under the FOSSology organization.

## I. INTRODUCTION

Text extraction is a very useful technology that has been growing with numerous additions and deletions in the approaches. Text extraction has its own meaning when we use it to extract the knowledge about the software out of the source codes. Source code these days contains a variety of knowledge like copyrights, license-text, docstrings (for documentation generation) and other metadata. All these forms of knowledge and information about the software are mentioned in the rich text of the source code file in the form of comments. Every programming language has a specific comment format and that is used to mention the particularity in the source code.

The Open Source world has a great impact on license-texts and copyrights in the source code because the source code is openly available for everyone and anyone to look at and use. The licensing and providing copyrights in the code provides recognition to the maintainers and the creators of the project. It also helps in managing and regulating the different usage of the source code by users/developers and prevents individuals/organizations from using code without permissions. There are various organizations like OSI and SPDX which provide different approved license statements. There are several big organizations from around the globe that are supporting the open-source culture and thus it gets mandatory to have something in the source code which regulates the rights for the individual/organizations work.

### A. Need for the metadata extraction

Metadata/Comments in the source code are an essential for the understanding of software in order to develop and maintain it [3][2][5]. There are several Open Source software that works for license compliance and license recognition of software and the basic requirements for them is to take the extracted comment as to their input and then do processing over it (Rule-based on machine learning). Apart from this, metadata contains some essential information about the software/source code which can be useful in various scenarios [7]. They help in improving the look and readability of the source code [3]. Commenting the snippet of the code by explaining it reduces the overall time and resources required for code maintenance [2][5][6]. The accurately classified and extracted input is very important [4][6]. If the input is noisy, the result score might get foggy and incorrect. If any of the machine learning approaches when used, a noisy input will lead to lesser accuracy. The noise here can be understood with the extraction of unwanted strings and code out from the software files.

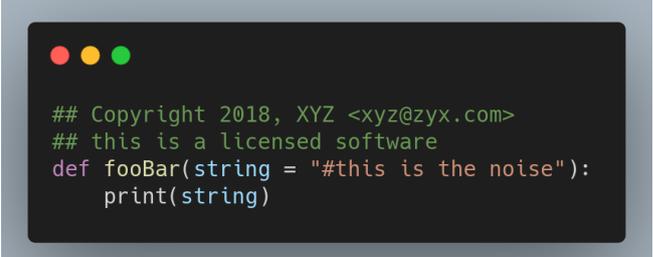

*Figure 1 - Example of a valid comment and noise*

In normal scenarios the proper extraction should be the first two lines of comments whereas, the string passed in the function is the unwanted noise which will later result in the reduced accuracy.

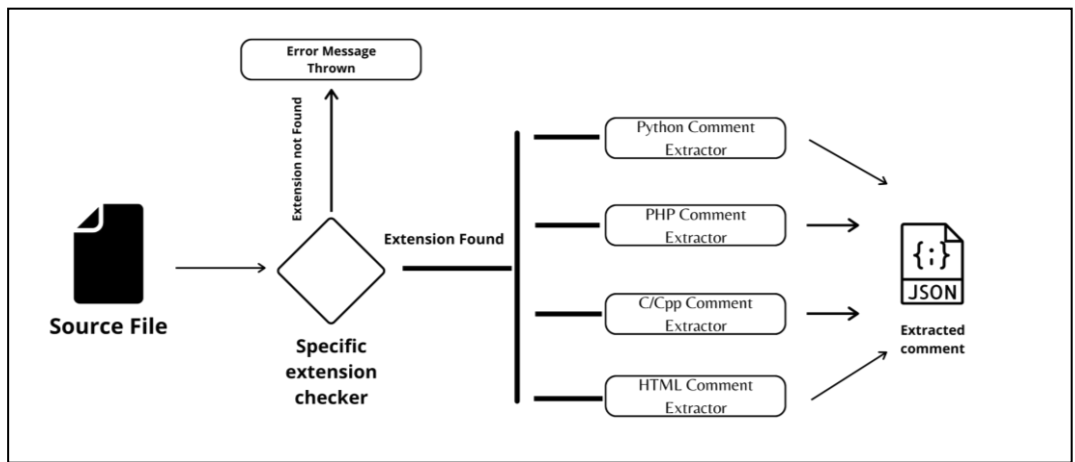

Figure 2 - Architecture of Nirjas

## II. RELATED WORK

Daniela et al. in their paper [3] did the quality anlysis and assesments of the code comments. They went on explaining how previous appraoches for the quality assesment was not sufficient. They developed a model to access the quality and evaluated it by doing a survey among developers.

Fluri et al. in their paper [2] talked about the evolution of code comments with respect to the source code. They found that the trend for commenting the newly added code is decreasing. Most of the commented parts are the declaration of functions and classes. Their work showed that around 97% of the time, the comments are revised whenever there are code modifications.

A. Rachman et al. in their paper [5] shed some light on the use of Regex for the extraction of comments and functions. Their work should us a way for the faster approach to tackle the problem and they were able to reach an accuracy of 100% for the text extraction.

A. Bartoli et al. worked on describing a system for synthesizing the required regular expression automatically [1]. They explained that the texts usually are structred in a way that it follows a syntactical pattern and can be best extracted using regular expressions.

A. Rachman, N. F. Rozi and A. Faradisa stated that the development and maintainance of softwares are very complex as well as costly and thus engenders the need for proper documentation about the source code [6]. They worked on dividing the source code with comments and later used the comments seperately to better understand the code. They also used regular expressions which helped them attain 100% of precision and recall for their code comment extraction.

E. Wong, T. Liu and L. Tan proposed a method to generate code comments automatically [7]. They used NLP techniques to find similar code segments and mapped the comments from the existing code segments to the new ones. They also evaluated the quality of their generated code comments and found out that only 23.7% of them can be labelled as good for which they proposed various improvement approaches.

P. Wang et al. in their paper [10], talked about how the use of regular expression evolved with time. They examined the entire commit history of the project and observed these changes.

## III. ARCHITECTURE

In the real-world scenario, mostly the metadata is found in source code collectively as in the form of comments which is a comprehension of the source code [9]. One of the approaches to do that was the Heuristic Regex based methodology where you have the essential detail of the comment syntax of the specific language and afterward utilizing that approach we can extricate every one of the comment out of the code. This approach was more ideal and accurate with the extraction because there was less noise which was extracted, and the high versatility of its working was even more effective.

Regex can be used to describe any given sequence or stream of texts [1]. The pattern seems convoluted at first but once understood correctly can be easily generated for any set of cases. It always provides great results with searches and validations and is often used for parsing and manipulating the set of characters or texts [5].

The approach used is Nirjas is more of an inclusion of techniques used in normal extraction of text from the source and a few upgrades for differentiating the metadata and the code methodology. This approach for extraction is further supported by Regex. A target source file/folder path is added to the command-line argument to run Nirjas on it. It then extracts all the source texts and applies the base algorithm to fetch us all the metadata.

### A. Comment Conventions

All programming languages have their own style and convention for comment or the metadata inside the source code. These strategies are then utilized with the help of regex to get an ideal extraction result.

Apart from different languages, Other things were that each programming language has its different kinds of comments as well i.e. Single Line comments, Multi-Line Comments and a combination of both single and multiline comments. Comments can be placed anywhere except for the code part but these insertions should be easily distinguishable [11].

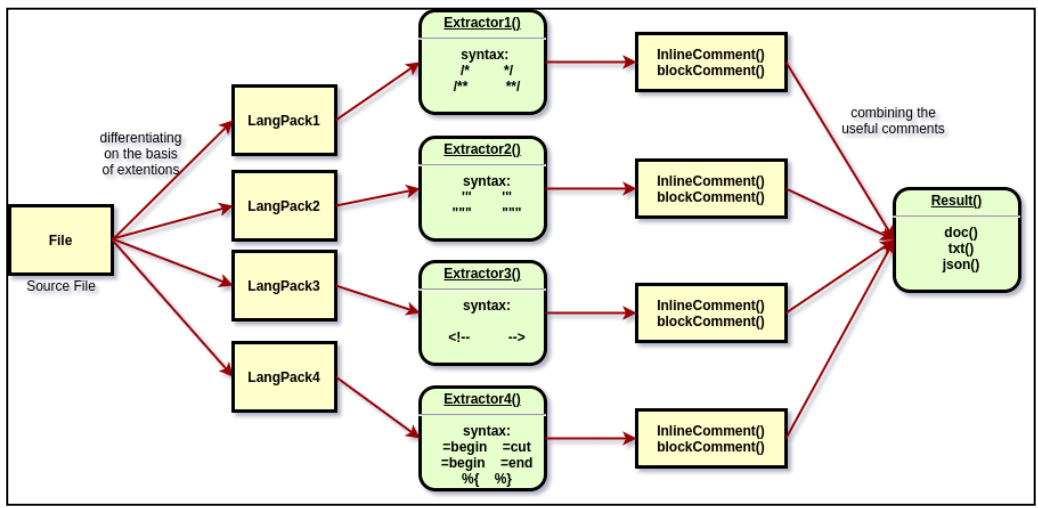

Figure 3 - Flow of comment-syntax mapping

Embedding comments deeply within the source code will make it less readable [3] and will hamper the original motivation for which the commenting process was introduced. Things like file description, copyright statements, license statements should be written in the first few lines of the source file. Developers do not comment often in the code [12] but this is observed that the commenting practice is done mostly in the beginning of the project. The developers do not generally bother adding/modifying comments whenever a new addition/modification is done [2]. Comments added near or on the same line as any function or class are usually the clarification of the respective function or class [2].

E.g. Comments in python file

```
# Single line comment #1
## Single line comment #2
''' Multiline comment: Line 1
Multiline comment: Line 2 '''

def fooBar(string = "#this is not a comment"):
    print(string)
```

Figure 4 - Comments in a python file

Line 1 & 2 are examples of syntax used for single line comment and Line 3 & 4 are examples of multi-line comments. This mixture of several types of comments into one single file takes us to devise a single process that can do the work for both single and multi-line embedded comments in a file.

## IV. ALGORITHM

Nirjas is a command line tool which takes the source file/folder as input and sequentially scans the rich text to differentiate between the source code and the metadata.

Nirjas is written purely in python and are supported for almost all the major programming language. The structural flow is defined in a way that it is very easy to provide support for any new programming language. The regular expression for almost all the commenting conventions are written and tested so that we do not miss out on any of the edge case.

The programming language of the input file is determined by stripping the extension of the file. Various extensions are mapped with their respective programming language. Once the language of the file is determined, the input is sent directly to its respective language function for further processing. Nirjas has separate file for all the supported languages which is kept so to reduce the effort in maintaining the existing one or adding a new one.

The language file fetches the syntax for different types of comments used by the particular language and uses those syntax to identify and organize the comments with its location. Nirjas also extracts various other information like SLOCs, blank lines, starting and ending line is the comment is multiline. These extracted information are further mapped within a single result that can be exported as JSON.

| Algorithm: High Level Pseudocode for Nirjas |
|---|
| **Require:** Input Path |
| **begin** |
| 1: **for** file **in** Input Path **do** |
| 2:    **Run** langIdentifier(file) |
| 3:    fileLang ← languageMapper[file extension] |
| 4:    **for** lang **in** fileLang **do** |
| 5:      readfile() |
| 6:      **fetch** Syntax ← CommentSyntax() |
| 7:      single_line ← result.singleLineSyntax(file) |
| 8:      multi_line ← result.multiLineSyntax(file) |
| 9:      **return** result ← map(single_line, multi_line) |
| 10:    **end for** |
| 11: **end for** |
| **end** |

## A. Lang Identifier and Regex extraction

The Nirjas's extension checker has a predefined data structure that contains the extension of all the supported languages i.e. C/CPP, C#, CSS, Dart, Go, Haskell, Python, R, Ruby and many more. The checker sends the source code to the specific comment style extractor and the function there returns the extracted comment in the form JSON containing, Single line comments, Multi-Line comments separately.

Thus, each programming language has two different regexes working for it or each language has different syntax working on it depending upon the amount of syntax it uses for commenting. It is a top down approach of recognizing the language from the extension and then sending the code to the specific language file and then sending it to a specific comment type function which then returns the dictionary of extracted comments out of code.

Table 1 - Languages with comment syntax and extraction rules

| Languages | Syntax | Regex/Extraction pattern |
|---|---|---|
| Perl, Python, R, Shell | # | (?<!["'`])#+\s*(.*) |
| MATLAB | % | (?<!["'`])\%\s*(.*) |
| C, C#, C++, Go, Java, JavaScript, Kotlin, PHP, Rust, Scala, Swift, TypeScript | // | (?<![pst"'`]:)\/\/\s*(.*) |
| Python | ' ' | extract(self.start, self.end) |
| Python | " " | extract(self.start, self.end) |
| Haskell | -- | (?<!["'`])\-\-\s*(.*) |
| C, C#, C++, CSS, Dart, Go, HTML, Java, JavaScript, Kotlin, PHP, Rust, Scala, SCSS, Swift, TypeScript | /* */ | extract(self.start, self.end) |
| HTML | <!-- --> | extract(self.start, self.end) |
| Perl | =begin =cut | extract(self.start, self.end) |
| Ruby | =begin =end | extract(self.start, self.end) |
| Haskell | {- -} | extract(self.start, self.end) |
| MATLAB | %{ %} | extract(self.start, self.end) |
| Dart, SCSS | /// | (?<!["'`])\/\/\/\s*(.*) |
| Dart, SCSS | // (no triple dash) | (?<!\/)\/\/(?!\/)\s*(.*) |
| Ruby | # (no Curl) | (?<!["'`])#+(?!\{)\s*(.*) |

Nirjas is designed to provide some relevant information about a piece of code other than the source code and these are:

- Different Types of comments used in the file
- Information containing the place and location where these comments exists
- Start line and end line for particular multiline comment
- Programming language of the file
- Total source line of code
- Total line of comments, code as well as blank lines
- Nirjas can also separate the pure source code out of the file

Regex was the ideal choice to use for extraction because it gives us the freedom to specify the matching technique and matching limit as well. For each programming language, the comment extraction regex is defined in the file itself. All the regexes extract full matching data out of the file, be it comments, inline comments etc. Regex has been seen to achieve 100% of the precision and recall [5][6] in its extraction once written perfectly.

## B. Other Approach: Non-Regex Method

The hard coding methodology was also tried in order to compare between regex and direct string-based extraction. The string-based extraction was working in a way that it will extract everything as soon as the matching symbol was encountered in it.

The problem with this approach was that a lot of noise during extraction was found. The output was not filtered and lots and lots of comments are missing. There are various edge cases where the metadata is embedded with the code base and it can be really challenging for the non-regex algorithms to extract those metadata.

## V. CHANLLENGES

The main challenge with the metadata extraction was to differentiate between the code and the comment part. We encountered a plethora of edge cases where the code was written using the characters which triggered the extraction algorithm to fetch it as a comment instead of a source code line. There are few cases where to make multiline comment, all the lines are commented using single comment line separately, but due to the continuity of lines it appears as a multiline comment. Generally, the comments are placed in the near vicinity of a function or class [2] which makes it really difficult to separate both the entities and extracting only the relevant one. This number is still less as it is observed that only 15.4% of the total functions or classes are commented [12].

Many a times, the comments were written so poorly that the algorithm captured a lot of extra noise with the comment. Extracting only the relevant part of the comment was also a big challenge as various special character were added inside the comment. We then studied the process through which the programming language's compiler uses to differentiate between a line of code and a comment which helped us to decide Regex as our main extraction agent.

We also encountered the cases where the code part was commented out, which is a common practice in development phase of a project. Since Nirjas does not perform semantical analysis of the text, it was impossible to map a commented part as a piece of code. Also, this practice is not appreciated in the production ready code, so we decided to neglect the edge case and label it as a comment.

## VI. RESULT AND CONCLUSION

After scanning the source code using Nirjas, the output result is stored in a separate file. The output format which aligns with regex was in the format of the JSON, aligning a separate output class for each extraction. The multi, single and the string extraction have their own keys and the out is taken from them.

Nirjas provides various options where we can extract comments from a single file, a folder containing sub folder and files inside them or a complete software/package source directory. Users has the freedom to specify whether they want the comments as the metadata or to extract the source code, which will exclude the comments and will give only the rich text.

The list of different comment/metadata that has been displayed in the output is explicitly determined utilizing a few strategies i.e. Total lines of code mean the total number of lines present in the code file. Total lines of comments mean the number of lines where comments are mentioned. There is also one feature for multi-line comments i.e. Start line: line from where the comment starts, End line: The line where the multi-line comment ends.

It is really time consuming and difficult to grasp the program file by reading the source code only. Nirjas helps us to extract all the different metadata and comments from the file which makes it easier to comprehend the program. Also, it is really easy for us to backtrack the specific implemented portion of the code if we have its descriptive comment handy. Nirjas also captures the position of the particular comment in the file which makes it easier to jump to the desired location.

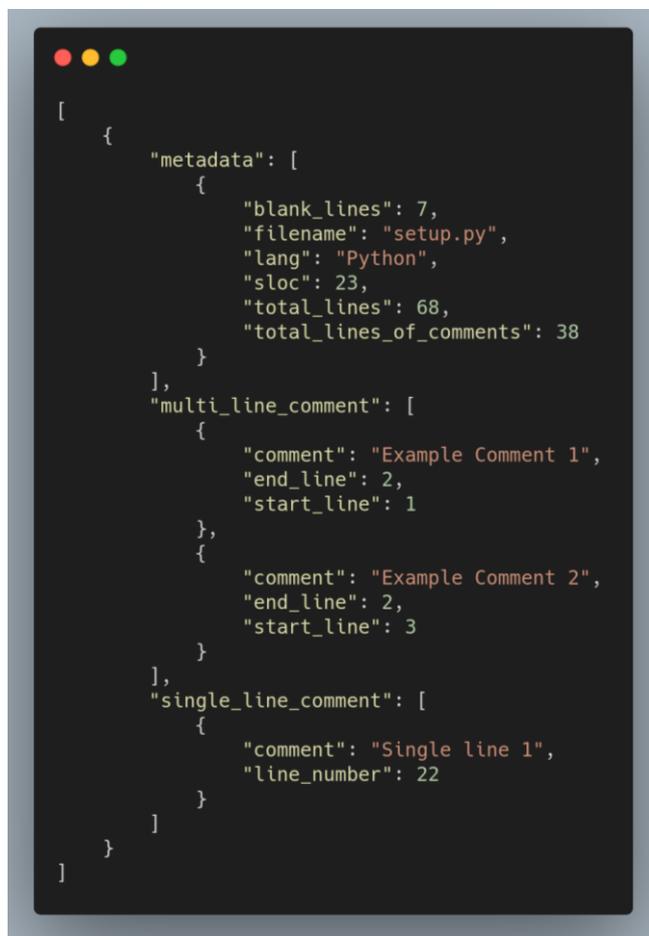

*Figure 5 - Example JSON result of Nirjas*

After extraction, each of the first lines is tracked and a line number is then noted. It acts as a check as soon as there is a match with the comment's regex, the line number is then updated, and other matrices are calculated using these core methodologies. There is an extension to other functionalities as well. The source code extraction: The SLOC has been calculated on the basis of what lines are actually regex and what is not and according to another feature of extracting not only the comments, but the source code was also introduced.

## VII. AVAILABILITY

The source code of Nirjas is publicly available on our GitHub [8] and is continuously developed and maintained under FOSSology Organization as its project. It contains the extensive user as well as developer documentation containing all the installation procedure. Nirjas is a standalone python library and can be separately installed using pip. The documentation contains most of the commands and its example usage which can be referred through its GitHub's wiki page [8].

## VIII. ACKNOWLEDGEMENT

We would like to thank the FOSSology community and The Linux Foundation for their continuous motivation and extensive support. Nirjas was first created and started out as a Google Summer of Code project in 2021, so we would like to extend our gratitude to Google for their GSoC program.